\documentclass[12pt,a4paper]{article}
\makeatletter
\def\eqnarray{%
\stepcounter{equation}%
\let\@currentlabel=\theequation
\global\@eqnswtrue
\global\@eqcnt\z@
\tabskip\@centering
\let\\=\@eqncr
$$\halign to \displaywidth\bgroup\@eqnsel\hskip\@centering
$\displaystyle\tabskip\z@{##}$&\global\@eqcnt\@ne
\hfil$\displaystyle{{}##{}}$\hfil
&\global\@eqcnt\tw@$\displaystyle\tabskip\z@{##}$\hfil
\tabskip\@centering&\llap{##}\tabskip\z@\cr}
\makeatother
\newcommand{\ket}[1]{{\vert{#1}\rangle}}
\newcommand{\bra}[1]{{\langle{#1}\vert}}
\newcommand{\braket}[1]{{\langle{#1}\vert{#1}\rangle}}

\newcommand{\fukuso}{{\mathbf C}}
\newcommand{\futon}{{\mathbf N}}

\begin{document}

\title{\slshape Jarlskog's Parametrization of Unitary Matrices and Qudit Theory}
\author{
  Kazuyuki FUJII
  \thanks{E-mail address : fujii@yokohama-cu.ac.jp }\ \ ,\ \ 
  Kunio FUNAHASHI
  \thanks{E-mail address : funahasi@isc.meiji.ac.jp }\quad and\ \ 
  Takayuki KOBAYASHI
  \thanks{E-mail address : s045559d@yokohama-cu.ac.jp }\\
  ${}^{*,\ \ddagger}$Department of Mathematical Sciences\\
  Yokohama City University\\
  Yokohama, 236--0027\\
  Japan\\
  ${}^{\dagger}$Division of Natural Science, Izumi Campus\\
  Meiji University\\
  Tokyo, 168--8555\\ 
  Japan
  }
\date{}
\maketitle
%
%
%
%
\begin{abstract}
  In the paper (math--ph/0504049) Jarlskog gave an interesting simple 
  parametrization to unitary matrices, which was essentially the canonical 
  coordinate of the second kind in the Lie group theory (math--ph/0505047). 
  
  In this paper we apply the method to a quantum computation based on 
  multi--level system (qudit theory). Namely, by considering that 
  the parametrization gives a complete set of modules in qudit theory, 
  we construct the generalized Pauli matrices which play a central role in  
  the theory and also make a comment on the exchange gate of two--qudit 
  systems. 
  
  Moreover we give an explicit construction to the generalized 
  Walsh--Hadamard matrix in the case of $n=3$, $4$ and $5$. 
  For the case of $n=5$ its calculation is relatively complicated.
  In general, a calculation to construct it tends to become more and more 
  complicated as $n$ becomes large. 

  To perform a quantum computation the generalized Walsh--Hadamard matrix  
  must be constructed in a quick and clean manner. From our construction 
  it may be possible to say that a qudit theory with $n \geq 5$ is not 
  realistic. 
  
  This paper is an introduction towards Quantum Engineering.
\end{abstract}
%


%
%
%
%

\section{Introduction}

In the paper \cite{CJ} Jarlskog gave a recursive parametrization to 
unitary matrices. See also \cite{Dita} as a similar parametrization. 
One of the authors showed that the recursive method was essentially obtained 
by the so--called canonical coordinate of the second kind in the Lie group 
theory, \cite{KF0}.

We are working in Quantum Computation, therefore we are interested in 
some application to quantum computation. 

One of key points of quantum computation is to construct some unitary 
matrices (quantum logic gates) in an efficient manner like Discrete 
Fourier Transform (DFT) when $n$ is large enough, \cite{PWS}. 
However, such a quick construction is in general not easy, see \cite{nine}  
or \cite{KF1}, \cite{KF2}.

The parametrization of unitary matrices given by Jarlskog may be convenient 
for our real purpose. 
We want to apply the method to quantum computation based on multi--level 
system (qudit theory).  One of reasons to study qudit theory comes from 
a deep problem on decoherence (we don't repeat the reason here). 
In the following let us consider an $n$ level system (for example, an atom 
has $n$ energy levels). 

Concerning an explicit construction of quantum logic gates in qudit theory, 
see for example \cite{KF3}, \cite{FHKW} and \cite{KuF}. 
By use of the new parametrization to unitary matrices we want to construct 
important logic gates {\bf in an explicit manner}\footnote{Quantum 
computation is not a pure mathematics, so we need an explicit construction}, 
especially the generalized Pauli matrices and Walsh--Hadamard matrix, which 
play a central role in qudit theory.

In this paper we construct the generalized Pauli matrices in a complete 
manner, while the Walsh--Hadamard matrix is constructed only for the case of 
$3$, $4$ and $5$ level systems. A calculation to construct it for the $5$ 
level system is relatively complicated compared to the $3$ and $4$ level 
systems.
In general, a calculation tends to become more and more complicated as $n$ 
becomes large. 

The generalized Walsh--Hadamard matrix gives a superposition of states in 
qudit theory, which is the heart of quantum computation.
It is natural for us to request {\bf a quick and clean construction} to it.

Therefore our calculation (or construction) may imply that a qudit theory 
with $n \geq 5$ is not realistic. Further study will be required.

\section{Jarlskog's Parametrization}

Let us make a brief introduction to the parametrization of unitary matrices 
by Jarlskog with the method developed in \cite{KF0}. 
The unitary group is defined as 
\begin{equation}
U(n)=\left\{\,U \in M(n,\fukuso)\ |\ U^{\dagger}U=UU^{\dagger}={\bf 1}_{n}\,\right\}
\end{equation}
and its (unitary) algebra is given by
\begin{equation}
u(n)=\left\{\,X \in M(n,\fukuso)\ |\ X^{\dagger}=-X\,\right\}.
\end{equation}
Then the exponential map is
\begin{equation}
\mathrm{exp} : u(n)\ \longrightarrow\ U(n)\ ;\ X\ \mapsto\ U\equiv \mathrm{e}^{X}.
\end{equation}
This map is canonical but not easy to calculate.

We write down the element $X \in u(n)$ explicitly :
\begin{equation}
X=
\left(
\begin{array}{cccccc}
i\theta_{1} & z_{12} & z_{13} & \cdots & z_{1,n-1} & z_{1n} \\
-\bar{z}_{12} & i\theta_{2} & z_{23} & \cdots & z_{2,n-1} & z_{2n} \\
-\bar{z}_{13} & -\bar{z}_{23} & i\theta_{3} & \cdots & z_{3,n-1} & z_{3n} \\
 \vdots & \vdots & \vdots & \ddots & \vdots & \vdots  \\
-\bar{z}_{1,n-1} & -\bar{z}_{2,n-1} & -\bar{z}_{3,n-1} & \cdots & 
i\theta_{n-1} & z_{n-1,n} \\
-\bar{z}_{1,n} & -\bar{z}_{2,n} & -\bar{z}_{3,n} & \cdots & -\bar{z}_{n-1,n}
& i\theta_{n}
\end{array}
\right).        
\end{equation}
This $X$ is decomposed into
\[
X=X_{0}+X_{2}+\cdots + X_{j}+\cdots +X_{n}
\]
where
\begin{equation}
X_{0}=
\left(
\begin{array}{cccccc}
i\theta_{1} & & & & &   \\
& i\theta_{2} & & & &   \\
& & i\theta_{3} & & &   \\
& & & \ddots &  &       \\
& & & & i\theta_{n-1} & \\
& & & & & i\theta_{n} 
\end{array}
\right)
\end{equation}
and for $2\leq j\leq n$
\begin{equation}
X_{j}=
\left(
\begin{array}{cccccc}
0 & & & & &   \\
& \ddots & & \ket{z_{j}} & &   \\
& & \ddots & & &   \\
& -\bra{z_{j}} &  & 0 &  &  \\
& & & & \ddots & \\
& & & & & 0
\end{array}
\right),\quad 
\ket{z_{j}}=
\left(
\begin{array}{c}
z_{1j} \\
z_{2j} \\
\vdots \\
z_{j-1,j}
\end{array}
\right).
\end{equation}

Then the canonical coordinate of the second kind in the unitary group (Lie 
group) is well--known and given by
\begin{equation}
\label{eq:the canonical coordinate of the second kind}
u(n) \ni X=X_{0}+X_{2}+\cdots + X_{j}+\cdots +X_{n}
\ \longrightarrow\ 
\mathrm{e}^{X_{0}}\mathrm{e}^{X_{2}}\cdots \mathrm{e}^{X_{j}}\cdots \mathrm{e}^{X_{n}}
 \in U(n)
\end{equation}
in this case \footnote{There are of course some variations}. Therefore 
we have only to calculate $\mathrm{e}^{X_{j}}$ for $j\geq 2$ ($j=0$ is trivial), 
which is easy (see Appendix). The result is
\begin{equation}
\label{eq:fundamental}
\mathrm{e}^{X_{j}}
=
\left(
\begin{array}{ccc}
{\bf 1}_{j-1}-\left(1-\cos(\sqrt{\braket{z_{j}}})\right)
\ket{\tilde{z}_{j}}\bra{\tilde{z}_{j}}& 
\sin(\sqrt{\braket{z_{j}}})\ket{\tilde{z}_{j}} &  \\
-\sin(\sqrt{\braket{z_{j}}})\bra{\tilde{z}_{j}} & 
\cos(\sqrt{\braket{z_{j}}}) & \\
  &  &  {\bf 1}_{n-j}
\end{array}
\right)
\end{equation}
where $\ket{\tilde{z}_{j}}$ is a normalized vector defined by
\begin{equation}
\ket{\tilde{z}_{j}}\equiv \frac{1}{\sqrt{\braket{z_{j}}}}\ket{z_{j}}
\ \Longrightarrow\ \braket{\tilde{z}_{j}}=1.
\end{equation}

\par \vspace{3mm}
We make a comment on the case of $n=2$. Since 
\[
\ket{\tilde{z}}=\frac{z}{|z|}\equiv \mathrm{e}^{i\alpha},\quad
\bra{\tilde{z}}=\frac{\bar{z}}{|z|}=\mathrm{e}^{-i\alpha}
\ \Longrightarrow\
\ket{\tilde{z}}\bra{\tilde{z}}=\braket{\tilde{z}}=1,
\]
we have
\begin{eqnarray}
\label{eq:Euler angle parametrization}
\mathrm{e}^{X_{0}}\mathrm{e}^{X_{2}}
&=&\left(
\begin{array}{cc}
\mathrm{e}^{i\theta_{1}} &        \\
   & \mathrm{e}^{i\theta_{2}}
\end{array}
\right)
\left(
\begin{array}{cc}
\cos(|z|) & \mathrm{e}^{i\alpha}\sin(|z|)    \\
-\mathrm{e}^{-i\alpha}\sin(|z|) & \cos(|z|)
\end{array}
\right)        \nonumber \\
&=&
\left(
\begin{array}{cc}
\mathrm{e}^{i\theta_{1}} &        \\
   & \mathrm{e}^{i\theta_{2}}
\end{array}
\right)
\left(
\begin{array}{cc}
\mathrm{e}^{i\alpha/2} &        \\
   & \mathrm{e}^{-i\alpha/2}
\end{array}
\right)
\left(
\begin{array}{cc}
\cos(|z|)  & \sin(|z|)    \\
-\sin(|z|) & \cos(|z|)
\end{array}
\right)
\left(
\begin{array}{cc}
\mathrm{e}^{-i\alpha/2} &        \\
   & \mathrm{e}^{i\alpha/2}
\end{array}
\right)    \nonumber \\
&=&
\left(
\begin{array}{cc}
\mathrm{e}^{i(\theta_{1}+\alpha/2)} &        \\
   & \mathrm{e}^{i(\theta_{2}-\alpha/2)}
\end{array}
\right)
\left(
\begin{array}{cc}
\cos(|z|)  & \sin(|z|)    \\
-\sin(|z|) & \cos(|z|)
\end{array}
\right)
\left(
\begin{array}{cc}
\mathrm{e}^{-i\alpha/2} &        \\
   & \mathrm{e}^{i\alpha/2}
\end{array}
\right).
\end{eqnarray}
This is just the Euler angle parametrization.

Therefore the parametrization (\ref{eq:the canonical coordinate of the second 
kind}) may be considered as a kind of generalization of Euler's angle one.

\par \vspace{5mm}
In the following we set 
\begin{eqnarray}
\label{eq:fundamental-modify}
A_{0}\equiv \mathrm{e}^{X_{0}}
&=&
\left(
\begin{array}{ccc}
\mathrm{e}^{i\theta_{1}} &        &                        \\
                       & \ddots &                        \\
                       &        & \mathrm{e}^{i\theta_{n}}
\end{array}
\right),     \nonumber \\
A_{j}\equiv \mathrm{e}^{X_{j}}
&=&
\left(
\begin{array}{ccc}
{\bf 1}_{j-1}-\left(1-\cos{\beta_{j}}\right)
\ket{\tilde{z}_{j}}\bra{\tilde{z}_{j}}& 
\sin{\beta_{j}}\ket{\tilde{z}_{j}} &  \\
-\sin{\beta_{j}}\bra{\tilde{z}_{j}} & \cos{\beta_{j}} & \\
  &  &  {\bf 1}_{n-j}
\end{array}
\right)
\end{eqnarray}
for $2\leq j\leq n$. More precisely, we write
\begin{equation}
\label{eq:modules}
A_{0}=A_{0}(\{\theta_{1},\theta_{2},\cdots,\theta_{n}\}),\quad 
A_{j}=A_{j}(\{\tilde{z}_{1j},\tilde{z}_{2j},\cdots,\tilde{z}_{j-1,j}\};
\beta_{j})\quad \textrm{for}\quad j=2,\cdots,n
\end{equation}
including parameters which we can manipulate freely.

From now on we consider $A_{j}$ {\bf a kind of module} of qudit theory for 
$j=0,2,\cdots,n$,\ namely $\{A_{j}|\ j=0,2,\cdots,n\}$ becomes a complete set 
of modules. By combining them many times 
\footnote{we take no account of an ordering or a uniqueness of $\{A_{j}\}$ in 
the expression (\ref{eq:the canonical coordinate of the second kind})}
we construct important matrices in qudit theory {\bf in an explicit manner}.

\section{Qudit Theory}

Let us make a brief introduction to a qudit theory. The theory is based on 
an atom with $n$ energy levels $\{(\ket{k},E_{k})\ |\ 0 \leq k \leq n-1\}$ 
, see the figure 1.

\begin{figure}
\begin{center}
\setlength{\unitlength}{1mm}   %
\begin{picture}(100,100)(0,0)
\put(0,80){\makebox(15,10)[c]{$E_{n-1}$}}
\put(15,85){\line(1,0){70}}
\put(85,80){\makebox(18,10)[c]{$|{n-1}\rangle$}}
\put(0,70){\makebox(15,10)[c]{$E_{n-2}$}}
\put(15,75){\line(1,0){70}}
\put(85,70){\makebox(18,10)[c]{$|{n-2}\rangle$}}
\put(5,50){\makebox(10,10)[c]{$E_2$}}
\put(15,55){\line(1,0){70}}
\put(85,50){\makebox(10,10)[c]{$|2\rangle$}}
\put(5,35){\makebox(10,10)[c]{$E_1$}}
\put(15,40){\line(1,0){70}}
\put(85,35){\makebox(10,10)[c]{$|1\rangle$}}
\put(5,15){\makebox(10,10)[c]{$E_0$}}
\put(15,20){\line(1,0){70}}
\put(85,15){\makebox(10,10)[c]{$|0\rangle$}}
\put(50,10){\circle*{3}}
\put(50,60){$\cdot$}
\put(50,65){$\cdot$}
\put(50,70){$\cdot$}
\put(50,30){\vector(0,1){10}}
\put(50,30){\vector(0,-1){10}}
\put(53,25){\makebox(10,10)[c]{$\omega_1$}}
\put(50,50){\vector(0,1){5}}
\put(50,50){\vector(0,-1){10}}
\put(53,42){\makebox(10,10)[c]{$\omega_2$}}
\put(50,80){\vector(0,1){5}}
\put(50,80){\vector(0,-1){5}}
\put(53,75){\makebox(10,10)[c]{$\omega_{n-1}$}}
\end{picture}
\vspace{-10mm}
\caption{Atom with $n$ energy levels}
\end{center}
\end{figure}

First of all we summarize some properties of the Pauli matrices and 
Walsh--Hadamard matrix, and next state corresponding ones of the generalized 
Pauli matrices and generalized Walsh--Hadamard matrix within our necessity. 

Let $\{\sigma_{1}, \sigma_{2}, \sigma_{3}\}$ be Pauli matrices : 
\begin{equation}
\label{eq:pauli}
\sigma_{1} = 
\left(
  \begin{array}{cc}
    0& 1 \\
    1& 0
  \end{array}
\right), \quad 
\sigma_{2} = 
\left(
  \begin{array}{cc}
    0& -i \\
    i& 0
  \end{array}
\right), \quad 
\sigma_{3} = 
\left(
  \begin{array}{cc}
    1& 0 \\
    0& -1
  \end{array}
\right).
\end{equation}
By (\ref{eq:pauli}) $\sigma_{2}=i\sigma_{1}\sigma_{3}$, so that the essential 
elements of Pauli matrices are $\{\sigma_{1}, \sigma_{3}\}$ and they satisfy
\begin{equation}
\sigma_{1}^{2}=\sigma_{3}^{2}={\bf 1}_{2}\ ;\quad 
\sigma_{1}^{\dagger}=\sigma_{1},\
\sigma_{3}^{\dagger}=\sigma_{3}\ ;\quad 
\sigma_{3}\sigma_{1}=-\sigma_{1}\sigma_{3}=\mathrm{e}^{i\pi}\sigma_{1}\sigma_{3}.
\end{equation}

A Walsh--Hadamard matrix is defined by 
\begin{equation}
   \label{eq:w-a}
   W = \frac{1}{\sqrt{2}}
     \left(
        \begin{array}{rr}
            1& 1 \\
            1& -1
        \end{array}
     \right)\ \in \ O(2)\ \subset U(2).
\end{equation}
This matrix is unitary and it plays a very important role 
in Quantum Computation. Moreover it is easy to realize it in Quantum Optics 
as shown in \cite{KF3}. 
Let us list some important properties of $W$ :
\begin{eqnarray}
\label{eq:properties of W-H (1)}
      &&W^{2}={\bf 1}_{2},\ \ W^{\dagger}=W=W^{-1}, \\
\label{eq:properties of W-H (2)}
      &&\sigma_{1}= W\sigma_{3}W^{-1},
\end{eqnarray}
The proof is very easy. 

Next let us generalize the Pauli matrices to higher dimensional cases. 
Let $\{\Sigma_{1}, \Sigma_{3}\}$ be the following matrices in $M(n,\fukuso)$
\begin{equation}
\label{eq:gener-pauli}
\Sigma_{1}=
\left(
\begin{array}{cccccc}
0&  &  &      &      &       1   \\
1& 0&  &      &      &           \\
  & 1& 0&      &      &          \\
  &  & 1& \cdot&      &          \\
  &  &  & \cdot& \cdot&          \\
  &  &  &      &    1 & 0
\end{array}
\right),      \qquad
\Sigma_{3}=
\left(
\begin{array}{cccccc}  
1&        &           &      &      &                  \\
  & \sigma&           &      &      &                  \\
  &       & {\sigma}^2&      &      &                  \\
  &       &           & \cdot&      &                  \\
  &       &           &      & \cdot&                  \\
  &       &           &      &      &  {\sigma}^{n-1}
\end{array}
\right)
\end{equation}
where $\sigma$ is a primitive root of unity ${\sigma}^{n}=1$ 
($\sigma=\mathrm{e}^{\frac{2\pi i}{n}}$). We note that
\[
\bar{\sigma}=\sigma^{n-1},\quad
1+\sigma+\cdots+\sigma^{n-1}=0 .
\]
Two matrices
$\{\Sigma_{1}, \Sigma_{3}\}$ are generalizations of the Pauli matrices
$\{\sigma_{1}, \sigma_{3}\}$, but they are not hermitian.
Here we list some of their important properties :
\begin{equation}
\Sigma_{1}^{n}=\Sigma_{3}^{n}={\bf 1}_{n}\ ; \quad
\Sigma_{1}^{\dagger}=\Sigma_{1}^{n-1},\
\Sigma_{3}^{\dagger}=\Sigma_{3}^{n-1}\ ; \quad
\Sigma_{3}\Sigma_{1}=\sigma \Sigma_{1}\Sigma_{3}\ .
\end{equation}
For $n=3$ and $n=4$ $\Sigma_{1}$ and its powers are given respectively as 
\begin{equation}
\label{eq:sigma-1-three}
\Sigma_{1}=
\left(
\begin{array}{ccc}
     0 &   & 1   \\
     1 & 0 &     \\
       & 1 & 0
\end{array}
\right),\quad 
\Sigma_{1}^{2}=
\left(
\begin{array}{ccc}
     0 & 1 &     \\
       & 0 & 1   \\
     1 &   & 0
\end{array}
\right)       
\end{equation}
and
\begin{equation}
\label{eq:sigma-1-four}
\Sigma_{1}=
\left(
\begin{array}{cccc}
   0 &   &    & 1  \\
   1 & 0 &    &    \\
     & 1 & 0  &    \\
     &   & 1  & 0
\end{array}
\right),\quad 
\Sigma_{1}^{2}=
\left(
\begin{array}{cccc}
   0 &   & 1  &    \\
     & 0 &    & 1  \\
   1 &   & 0  &    \\
     & 1 &    & 0
\end{array}
\right),\quad 
\Sigma_{1}^{3}=
\left(
\begin{array}{cccc}
   0 & 1 &    &    \\
     & 0 & 1  &    \\
     &   & 0  & 1  \\
   1 &   &    & 0
\end{array}
\right).
\end{equation}

If we define a Vandermonde matrix $W$ based on $\sigma$ as
\begin{eqnarray}
\label{eq:Large-double}
W&=&\frac{1}{\sqrt{n}}
\left(
\begin{array}{ccccccc}
1&        1&     1&   \cdot & \cdot  & \cdot & 1             \\
1& \sigma^{n-1}& \sigma^{2(n-1)}&  \cdot& \cdot& \cdot& \sigma^{(n-1)^2} \\
1& \sigma^{n-2}& \sigma^{2(n-2)}&  \cdot& \cdot& \cdot& \sigma^{(n-1)(n-2)} \\
\cdot&  \cdot &  \cdot  &     &      &      & \cdot  \\
\cdot&  \cdot  & \cdot &      &      &      &  \cdot  \\
1& \sigma^{2}& \sigma^{4}& \cdot & \cdot & \cdot & \sigma^{2(n-1)} \\
1& \sigma & \sigma^{2}& \cdot& \cdot& \cdot& \sigma^{n-1}
\end{array}
\right), \\
\label{eq:Large-double-dagger}
W^{\dagger}&=&\frac{1}{\sqrt{n}}
\left(
\begin{array}{ccccccc}
1&        1&     1&   \cdot & \cdot  & \cdot & 1             \\
1& \sigma& \sigma^{2}&  \cdot& \cdot& \cdot& \sigma^{n-1} \\
1& \sigma^{2}& \sigma^{4}&  \cdot& \cdot& \cdot& \sigma^{2(n-1)} \\
\cdot&  \cdot &  \cdot  &     &      &      & \cdot  \\
\cdot&  \cdot  & \cdot &      &      &      &  \cdot  \\
1& \sigma^{n-2}& \sigma^{2(n-2)}& \cdot& \cdot& \cdot& \sigma^{(n-1)(n-2)} \\
1&    \sigma^{n-1} & \sigma^{2(n-1)}& \cdot& \cdot& \cdot& \sigma^{(n-1)^2}
\end{array}
\right),
\end{eqnarray}
then it is not difficult to see
\begin{eqnarray}
\label{eq:properties of G-W-H (1)}
&&W^{\dagger}W=WW^{\dagger}={\bf 1}_{n},  \\
\label{eq:properties of G-W-H (2)}
&&\Sigma_{1}=W\Sigma_{3}W^{\dagger}=W\Sigma_{3}W^{-1}.
\end{eqnarray}

Since $W$ corresponds to the Walsh--Hadamard matrix (\ref{eq:w-a}), 
it may be possible to call $W$ the generalized Walsh--Hadamard matrix. 
If we write $W^{\dagger}=(w_{ab})$, then 
\[
w_{ab}=\frac{1}{\sqrt{n}}\sigma^{ab}=
\frac{1}{\sqrt{n}}\mathrm{exp}\left(\frac{2\pi i}{n}ab\right) 
\quad \textrm{for}\quad 0\leq a,\ b \leq n-1. 
\]
This is just the coefficient matrix of Discrete Fourier Transform (DFT) 
if $n=2^{k}$ for some $k \in \futon$, see \cite{PWS}. 

For $n=3$ and $n=4$ $W$ is given respectively as 
\begin{equation}
\label{eq:w-a-three}
W=
\frac{1}{\sqrt{3}}
\left(
\begin{array}{ccc}
1&   1&   1              \\
1& \sigma^{2} & \sigma   \\
1& \sigma & \sigma^{2}
\end{array}
\right)
=
\frac{1}{\sqrt{3}}
\left(
\begin{array}{ccc}
1&          1&                     1                 \\
1& \frac{-1-i\sqrt{3}}{2} & \frac{-1+i\sqrt{3}}{2}   \\
1& \frac{-1+i\sqrt{3}}{2} & \frac{-1-i\sqrt{3}}{2}
\end{array}
\right)
\end{equation}
and 
\begin{equation}
\label{eq:w-a-four}
W=
\frac{1}{2}
\left(
\begin{array}{cccc}
 1 &  1 &  1 &  1                       \\
 1 & \sigma^{3} & \sigma^{2} & \sigma   \\
 1 & \sigma^{2} & 1 & \sigma^{2}        \\
 1 & \sigma & \sigma^{2} & \sigma^{3}
\end{array}
\right)
=
\frac{1}{2}
\left(
\begin{array}{cccc}
 1 &  1 &  1 &  1   \\
 1 & -i & -1 &  i   \\
 1 & -1 &  1 & -1   \\
 1 &  i & -1 & -i
\end{array}
\right).
\end{equation}

\vspace{5mm}
We note that the generalized Pauli and Walsh--Hadamard matrices in three and 
four level systems can be constructed in a quantum optical manner (by using 
Rabi oscillations of several types), see \cite{KF3} and \cite{FHKW}. 
Concerning an interesting application of the generalized Walsh--Hadamard one 
in three and four level systems to algebraic equation, see \cite{KF4}.

\section{Explicit Construction of the Generalized Pauli and \\
Generalized Walsh--Hadamard Matrices}

First let us construct the generalized Pauli matrices. 
From (\ref{eq:modules}) it is easy to see
\begin{equation}
\label{eq:Sigma-3}
A_{0}(\{0,2\pi/n,4\pi/n,\cdots,2(n-1)\pi/n\})=\Sigma_{3}.
\end{equation}
Next we construct $\Sigma_{1}$. From (\ref{eq:modules}) we also set
\begin{equation}
A_{j}=A_{j}(\{0,\cdots,0,1\};\pi/2)=
\left(
\begin{array}{cccc}
{\bf 1}_{j-2} &    &   &                \\
              & 0  & 1 &                \\
              & -1 & 0 &                \\
              &    &   & {\bf 1}_{n-j}
\end{array}
\right)
\end{equation}
for $j=2,\cdots,n$. Then it is not difficult to see
\begin{equation}
A_{2}A_{3}\cdots A_{n}
=
\left(
\begin{array}{ccccc}
0  &    &        &        & 1   \\
-1 & 0  &        &        &     \\
0  & -1 & 0      &        &     \\
   &    & \ddots & \ddots &     \\
   &    &        & -1     & 0 
\end{array}
\right).
\end{equation}
Therefore if we choose $A_{0}$ as
\begin{equation}
A_{0}=A_{0}(\{0,\pi,\cdots,\pi\})
=
\left(
\begin{array}{ccccc}
1 &    &    &        &     \\
  & -1 &    &        &     \\
  &    & -1 &        &     \\
  &    &    & \ddots &     \\
  &    &    &        & -1
\end{array}
\right)
\end{equation}
then we finally obtain 
\begin{equation}
\label{eq:Sigma-1}
A_{0}A_{2}A_{3}\cdots A_{n}=\Sigma_{1}.
\end{equation}
From (\ref{eq:Sigma-3}) and (\ref{eq:Sigma-1}) we can construct all the 
generalized Pauli matrices \\
$\left\{\mathrm{e}^{i\phi(a,b)}\Sigma_{1}^{a}\Sigma_{3}^{b}\ |\ 0\leq a,\ b\leq 
n-1\right\}$, 
where $\mathrm{e}^{i\phi(a,b)}$ is a some phase depending on $a$ and $b$.

\vspace{3mm}
Similarly we can construct the matrix
\begin{equation}
K=
\left(
\begin{array}{cccccc}
1 &   &       &       &   &    \\
  &   &       &       &   & 1  \\
  &   &       &       & 1 &    \\
  &   &       & \cdot &   &    \\
  &   & \cdot &       &   &    \\
  & 1 &       &       &   &   
\end{array}
\right)
\end{equation}
as follows. If $n=2k$, then
\begin{eqnarray}
&&A_{0}(\{\underbrace{0,\cdots,0}_{\mbox{$k+1$}},
\underbrace{\pi,\cdots,\pi}_{\mbox{$k-1$}}\})
A_{k+2}(\{0,\cdots,0,1,0\};\pi/2)
A_{k+3}(\{0,\cdots,0,1,0,0\};\pi/2)
\cdots \nonumber \\
\times &&A_{2k-1}(\{0,0,1,0,\cdots,0\};\pi/2)
A_{2k}(\{0,1,0,\cdots,0\};\pi/2)
=K
\end{eqnarray}
and if $n=2k-1$, then
\begin{eqnarray}
&&A_{0}(\{\underbrace{0,\cdots,0}_{\mbox{$k$}},
\underbrace{\pi,\cdots,\pi}_{\mbox{$k-1$}}\})
A_{k+1}(\{0,\cdots,0,1\};\pi/2)
A_{k+2}(\{0,\cdots,0,1,0\};\pi/2)
\cdots \nonumber \\
\times &&A_{2k-2}(\{0,0,1,0,\cdots,0\};\pi/2)
A_{2k-1}(\{0,1,0,\cdots,0\};\pi/2)
=K.
\end{eqnarray}

Both $\Sigma_{1}$ and $K$ play an important role in constructing the exchange 
(swap) gate in two--qudit systems like the figure 2 where 
$\Sigma$=$\Sigma_{1}$. To be more precise, see \cite{KF5}.

\begin{figure}
\begin{center}
\setlength{\unitlength}{1mm}
\begin{picture}(200,35)
\put(25,30){\line(1,0){42}}  
\put(25,5){\line(1,0){10}}   
\put(41,5){\line(1,0){11}}   
\put(58,5){\line(1,0){27}}   
\put(55,8){\line(0,1){22}}   
\put(38,5){\circle{6}} 
\put(35,0){\makebox(6,10){K}} 
\put(14,25){\makebox(9,10)[r]{$|a\rangle$}} 
\put(14,0){\makebox(9,10)[r]{$|b\rangle$}} 
\put(52,25){\makebox(6,10){$\bullet$}} 
\put(55,5){\circle{6}} 
\put(52,0){\makebox(6,10){$\Sigma$}}
\put(70,30){\circle{6}}               
\put(67,25){\makebox(6,10){K}}  
\put(73,30){\line(1,0){9}}      
\put(85,5){\line(1,0){27}}   
\put(88,30){\line(1,0){9}}   
\put(103,30){\line(1,0){12}}  
\put(85,5){\line(0,1){22}}  
\put(82,0){\makebox(6,10){$\bullet$}} 
\put(85,30){\circle{6}}  
\put(82,25){\makebox(6,10){$\Sigma$}} 
\put(100,30){\circle{6}}  
\put(97,25){\makebox(6,10){K}}  
\put(115,30){\line(1,0){22}} 
\put(118,5){\line(1,0){19}} 
\put(115,8){\line(0,1){22}} 
\put(135,25){\makebox(9,10)[r]{$|b\rangle$}} 
\put(135,0){\makebox(9,10)[r]{$|a\rangle$}} 
\put(112,25){\makebox(6,10){$\bullet$}} 
\put(115,5){\circle{6}} 
\put(112,0){\makebox(6,10){$\Sigma$}}
\end{picture}
\caption{Exchange gate on two--qudit system}
\end{center}
\end{figure}
\par \noindent

It is interesting to note the simple relation
\begin{equation}
W^{2}=K.
\end{equation}
Namely, the generalized Walsh--Hadamard matrix $W$ (\ref{eq:Large-double}) is 
a square root of $K$.

\vspace{5mm}
Second we want to construct the generalized Walsh--Hadamard matrix, which is 
however very hard. Let us show only the case of $n=3$, $4$ and $5$. 
\begin{flushleft}
(a)\ $n$ = 3\ :\ For (\ref{eq:w-a-three}) we have
\end{flushleft} 
\begin{equation}
\label{eq:Walsh-Hadamard 3}
W=A_{0}A_{3}A_{2}A_{0}^{'}
\end{equation}
where each of matrices is given by
\begin{eqnarray*}
A_{0}&=&A_{0}(\{0,2\pi/3,4\pi/3\})=
\left(
\begin{array}{ccc}
1 &                    &                    \\
  & \mathrm{e}^{i2\pi/3} &                    \\
  &                    & \mathrm{e}^{i4\pi/3}
\end{array}
\right),   \\
A_{0}^{'}&=&A_{0}^{'}(\{-\pi/12,7\pi/12,0\})=
\left(
\begin{array}{ccc}
\mathrm{e}^{-i\pi/12} &                       &    \\
                      & \mathrm{e}^{i7\pi/12} &    \\
                      &                       & 1
\end{array}
\right)
\end{eqnarray*}
and
\[
A_{3}=A_{3}\left(\{1/\sqrt{2},1/\sqrt{2}\};\cos^{-1}(1/\sqrt{3})\right)
=\frac{1}{\sqrt{3}}
\left(
\begin{array}{rrr}
 \frac{\sqrt{3}+1}{2} & -\frac{\sqrt{3}-1}{2} & 1   \\
-\frac{\sqrt{3}-1}{2} &  \frac{\sqrt{3}+1}{2} & 1   \\
-1                    & -1                    & 1
\end{array}
\right)
\]
and
\[
A_{2}=A_{2}\left(\{\mathrm{e}^{-i\pi/2}\};\pi/4\right)
=
\left(
\begin{array}{rrr}
 \frac{1}{\sqrt{2}} & -\frac{i}{\sqrt{2}} &    \\
-\frac{i}{\sqrt{2}} &  \frac{1}{\sqrt{2}} &    \\
                    &                     & 1
\end{array}
\right).
\]
Here we have used 
\[
\cos(\pi/12)=\frac{\sqrt{6}+\sqrt{2}}{4}\quad \mbox{and}\quad
\sin(\pi/12)=\frac{\sqrt{6}-\sqrt{2}}{4}.
\]
In this case, the number of modules is $4$.

\begin{flushleft}
(b)\ $n$ = 4 :\ For (\ref{eq:w-a-four}) we have
\end{flushleft}
\begin{equation}
\label{eq:Walsh-Hadamard 4}
W=A_{0}A_{4}SA_{3}A_{2}A_{0}^{'}S
\end{equation}
where each of matrices is given by
\begin{eqnarray*}
A_{0}&=&A_{0}(\{0,2\pi/4,4\pi/4,6\pi/4\})=
\left(
\begin{array}{cccc}
1 &   &    &      \\
  & i &    &      \\
  &   & -1 &      \\
  &   &    & -i
\end{array}
\right),   \\
A_{0}^{'}&=&A_{0}(\{\pi/4,5\pi/4,0,0\})=
\left(
\begin{array}{cccc}
 \frac{1+i}{\sqrt{2}} &                       &   &    \\
                      & -\frac{1+i}{\sqrt{2}} &   &    \\
                      &                       & 1 &    \\
                      &                       &   & 1
\end{array}
\right)
\end{eqnarray*}
and
\[
A_{4}=A_{4}\left(\{1/\sqrt{3},1/\sqrt{3},1/\sqrt{3}\};\pi/3\right)=
\left(
\begin{array}{rrrr}
 \frac{5}{6} & -\frac{1}{6} & -\frac{1}{6} & \frac{1}{2}   \\
-\frac{1}{6} &  \frac{5}{6} & -\frac{1}{6} & \frac{1}{2}   \\
-\frac{1}{6} & -\frac{1}{6} &  \frac{5}{6} & \frac{1}{2}   \\
-\frac{1}{2} & -\frac{1}{2} & -\frac{1}{2} & \frac{1}{2}
\end{array}
\right)
\]
and
\[
A_{3}=A_{3}\left(\{1/\sqrt{2},1/\sqrt{2}\};\cos^{-1}(-1/3)\right)=
\left(
\begin{array}{rrrr}
 \frac{1}{3} & -\frac{2}{3} &  \frac{2}{3} &    \\
-\frac{2}{3} &  \frac{1}{3} &  \frac{2}{3} &    \\
-\frac{2}{3} & -\frac{2}{3} & -\frac{1}{3} &    \\
             &              &              & 1
\end{array}
\right)
\]
and
\[
S=A_{0}(\{0,0,\pi,0\})A_{3}(\{0,1\};\pi/2)=
\left(
\begin{array}{cccc}
1 &   &   &    \\
  & 0 & 1 &    \\
  & 1 & 0 &    \\
  &   &   & 1
\end{array}
\right)
\]
and
\[
A_{2}=A_{2}\left(\{\mathrm{e}^{i\pi/2}\};\pi/4\right)=
\left(
\begin{array}{cccc}
 \frac{1}{\sqrt{2}} & \frac{i}{\sqrt{2}} &   &    \\
 \frac{i}{\sqrt{2}} & \frac{1}{\sqrt{2}} &   &    \\
                    &                    & 1 &    \\
                    &                    &   & 1
\end{array}
\right).
\]
In this case, the number of modules is $9$.

Last we show a calculation for the case of $n=5$. However, it is 
relatively complicated as shown in the following. 
\begin{flushleft}
(c)\ $n$ = 5 :\ We have
\end{flushleft}
\begin{equation}
\label{eq:Walsh-Hadamard 5}
W=A_{0}A_{5}A_{4}S_{1}A_{3}S_{2}A_{0}^{'}
\end{equation}
where each of matrices is given by
\begin{eqnarray*}
A_{0}&=&A_{0}(\{0,2\pi/5,4\pi/5,6\pi/5,8\pi/5\})=
\left(
\begin{array}{ccccc}
1 &        &            &            &             \\
  & \sigma &            &            &             \\
  &        & \sigma^{2} &            &             \\
  &        &            & \sigma^{3} &             \\
  &        &            &            & \sigma^{4}
\end{array}
\right),      \\
A_{0}^{'}&=&A_{0}(\{9\pi/10,13\pi/10,-3\pi/10,\pi/10,0\})=
\left(
\begin{array}{ccccc}
\mathrm{e}^{i9\pi/10} &   &   &   &        \\
  & \mathrm{e}^{i13\pi/10} &  &   &        \\
  &  & \mathrm{e}^{-i3\pi/10} &   &        \\
  &  &  & \mathrm{e}^{i\pi/10} &           \\
  &     &            &            & 1
\end{array}
\right)       
\end{eqnarray*}
where 
\[
\sigma=\mathrm{e}^{i2\pi/5}=\cos(2\pi/5)+i\sin(2\pi/5)=
\frac{\sqrt{5}-1}{4}+i\frac{\sqrt{10+2\sqrt{5}}}{4}
\]
and
\begin{eqnarray*}
A_{5}
&=&A_{5}\left(\{1/2,1/2,1/2,1/2\};\cos^{-1}(1/\sqrt{5})\right) \nonumber \\
&=&
\frac{1}{\sqrt{5}}
\left(
\begin{array}{ccccc}
 \frac{3\sqrt{5}+1}{4} & -\frac{\sqrt{5}-1}{4} & -\frac{\sqrt{5}-1}{4} & 
 -\frac{\sqrt{5}-1}{4} & 1           \\
-\frac{\sqrt{5}-1}{4} & \frac{3\sqrt{5}+1}{4} & -\frac{\sqrt{5}-1}{4} & 
 -\frac{\sqrt{5}-1}{4} & 1          \\
-\frac{\sqrt{5}-1}{4} & -\frac{\sqrt{5}-1}{4} & \frac{3\sqrt{5}+1}{4} & 
 -\frac{\sqrt{5}-1}{4} & 1           \\
-\frac{\sqrt{5}-1}{4} & -\frac{\sqrt{5}-1}{4} & -\frac{\sqrt{5}-1}{4} &          \frac{3\sqrt{5}+1}{4} & 1          \\
-1 & -1 & -1& -1 & 1
\end{array}
\right)
\end{eqnarray*}
and
\[
A_{4}=A_{4}\left(\{a/u,\alpha/u,-\bar{\alpha}/u\};\theta_{4}\right)
=
\left(
\begin{array}{ccccc}
1-sa^{2} & -sa\bar{\alpha} & sa\alpha & \frac{a}{\sqrt{5}} &            \\
-sa\alpha & 1-s|\alpha|^{2} & s\alpha^{2} & \frac{\alpha}{\sqrt{5}} &   \\
sa\bar{\alpha} & s\bar{\alpha}^{2} & 1-s|\alpha|^{2} & 
-\frac{\bar{\alpha}}{\sqrt{5}} &                                        \\
-\frac{a}{\sqrt{5}} & -\frac{\bar{\alpha}}{\sqrt{5}} & 
\frac{\alpha}{\sqrt{5}} & -\frac{a}{\sqrt{5}} &                         \\
               &           &            &            & 1
\end{array}
\right)  
\]
where 
\begin{eqnarray*}
&&a\equiv \sin(2\pi/5),\quad 
\alpha\equiv \sin(\pi/5)+i\frac{\sqrt{5}}{2}=
\frac{\sqrt{10-2\sqrt{5}}}{4}+i\frac{\sqrt{5}}{2},\quad 
\cos(\theta_{4})\equiv -\frac{a}{\sqrt{5}}                \\
&&u\equiv \sqrt{4+\cos^{2}(2\pi/5)},\quad 
s\equiv 
\frac{2(35+\sqrt{5})}{305}\left(1+\frac{\sin(2\pi/5)}{\sqrt{5}}\right)
\end{eqnarray*}
and
\begin{eqnarray*}
S_{1}&=&A_{0}(\{0,\pi,\pi,0,0\})A_{2}(\{1\};\pi/2)A_{3}(\{0,1\};\pi/2)=
\left(
\begin{array}{ccccc}
0 &   & 1 &   &     \\
1 & 0 &   &   &     \\
  & 1 & 0 &   &     \\
  &   &   & 1 &     \\
  &   &   &   & 1
\end{array}
\right),   \\
S_{2}&=&A_{0}(\{0,0,\pi,0,0\})A_{3}(\{1,0\};\pi/2)=
\left(
\begin{array}{ccccc}
0 &   & 1 &   &     \\
  & 1 &   &   &     \\
1 &   & 0 &   &     \\
  &   &   & 1 &     \\
  &   &   &   & 1
\end{array}
\right)     
\end{eqnarray*}
and
\[
A_{3}=A_{3}\left(\{-\beta/v,\bar{\beta}/v\};\theta_{3}\right)
=
\left(
\begin{array}{ccccc}
1-\frac{|\beta|^{2}}{\sqrt{5}\left(\sqrt{5}-(a+ta^{2})\right)} & 
\frac{\beta^{2}}{\sqrt{5}\left(\sqrt{5}-(a+ta^{2})\right)} & 
-\frac{\beta}{\sqrt{5}} &   &        \\
\frac{\bar{\beta}^{2}}{\sqrt{5}\left(\sqrt{5}-(a+ta^{2})\right)} & 
1-\frac{|\beta|^{2}}{\sqrt{5}\left(\sqrt{5}-(a+ta^{2})\right)} & 
\frac{\bar{\beta}}{\sqrt{5}} &   &   \\
\frac{\bar{\beta}}{\sqrt{5}} & -\frac{\beta}{\sqrt{5}} & 
-\frac{a+ta^{2}}{\sqrt{5}} &  &      \\
  &  &  & 1 &    \\
  &  &  &   & 1
\end{array}
\right)     
\]
where 
\begin{eqnarray*}
&&\beta\equiv \bar{\alpha}+ta\alpha,\quad 
v\equiv \sqrt{5-(a+ta^{2})^{2}}=\sqrt{2}|\beta|,\quad 
\cos(\theta_{3})\equiv -\frac{a+ta^{2}}{\sqrt{5}}       \\
&&
t\equiv \frac{2(7\sqrt{5}+1)}{61}\left(1+\frac{\sin(2\pi/5)}{\sqrt{5}}\right).
\end{eqnarray*}

\vspace{5mm}
A comment is in order.

\par \noindent
(1) Our construction is not necessarily minimal, namely a number of 
modules can be reduced, see \cite{KuF2}.

\par \noindent
(2) For $n \geq  6$ we have not succeeded in obtaining the formula like 
(\ref{eq:Walsh-Hadamard 3}) or (\ref{eq:Walsh-Hadamard 4}) or 
(\ref{eq:Walsh-Hadamard 5}). 
In general, a calculation tends to become more and more complicated as $n$ 
becomes large. 

\par \noindent
(3) The heart of quantum computation is a superposition of (possible) states. 
Therefore a superposition like 
\[
\ket{0} \longrightarrow 
\frac{\ket{0}+\ket{1}+\ket{2}+\ket{3}+\ket{4}}{\sqrt{5}}
\]
in the $5$ level system must be constructed {\bf in a quick and clean manner}. 
The generalized Walsh--Hadamard matrix just gives such a superposition.  
Our construction in the system seems to be complicated, from which one may be 
able to conclude that a qudit theory with $n \geq 5$ is not realistic.

\section{Discussion}

In this paper we treated the Jarlskog's parametrization of unitary 
matrices as a complete set of modules in qudit theory and 
constructed the generalized Pauli matrices in the general case and 
generalized Walsh--Hadamard matrix in the case of $n=3$, $4$ and $5$. 

In spite of every efforts we could not construct the Walsh--Hadamard 
matrix in the general case, so its construction is left as a future task.

However, our view is negative to this problem. 
In general, a calculation to construct it tends to become more and more 
complicated as $n$ becomes large. See (\ref{eq:Walsh-Hadamard 3}) and  
(\ref{eq:Walsh-Hadamard 4}) and (\ref{eq:Walsh-Hadamard 5}). 
Therefore it may be possible to say that a qudit theory with $n \geq 5$  
is not realistic. Further study will be required.

Our next task is to realize these modules $\{A_{j}|\ j=0,2,\cdots,n\}$ 
in a quantum optical method, which will be discussed in a forthcoming paper.

\vspace{5mm}
\noindent{\em Acknowledgment.}\\
K. Fujii wishes to thank Mikio Nakahara and Shin'ichi Nojiri for their 
helpful comments and suggestions.

\vspace{5mm}
\begin{center}
 \begin{Large}
\noindent{\bfseries Appendix\quad Proof of the Formula (\ref{eq:fundamental})}
 \end{Large}
\end{center}
%

In this appendix we derive the formula (\ref{eq:fundamental}) to make 
the paper self--contained. 

\par \noindent
Since
\[
X_{j}=
\left(
\begin{array}{ccc}
{\bf 0}_{j-1} & \ket{z_{j}} &   \\
-\bra{z_{j}} & 0 &              \\
& & {\bf 0}_{n-j}
\end{array}
\right)
\equiv 
\left(
\begin{array}{cc}
K &               \\
  & {\bf 0}_{n-j}
\end{array}
\right)
\quad \Longrightarrow \quad
\mathrm{e}^{X_{j}}=
\left(
\begin{array}{cc}
 \mathrm{e}^{K} &     \\
   & {\bf 1}_{n-j}
\end{array}
\right)
\]
we have only to calculate the term $\mathrm{e}^{K}$, which is an easy task. 
From
\begin{eqnarray}
K&=&
\left(
\begin{array}{cc}
{\bf 0}_{j-1} & \ket{z_{j}} \\
-\bra{z_{j}}  & 0
\end{array}
\right),\quad 
K^{2}=
\left(
\begin{array}{cc}
-\ket{z_{j}}\bra{z_{j}} &  \\
  & -\braket{z_{j}}
\end{array}
\right),   \nonumber \\
K^{3}&=&
\left(
\begin{array}{cc}
{\bf 0}_{j-1} & -\braket{z_{j}}\ket{z_{j}}  \\
\braket{z_{j}}\bra{z_{j}} & 0            
\end{array}
\right)
=-\braket{z_{j}}K
\end{eqnarray}
we have important relations
\[
K^{2n+1}=\left(-\braket{z_{j}}\right)^{n}K,\quad 
K^{2n+2}=\left(-\braket{z_{j}}\right)^{n}K^{2}
\quad \mbox{for}\quad n\geq 0.
\]
Therefore
\begin{eqnarray*}
\mathrm{e}^{K}
&=&{\bf 1}_{j}+\sum_{n=0}^{\infty}\frac{1}{(2n+2)!}K^{2n+2}
              +\sum_{n=0}^{\infty}\frac{1}{(2n+1)!}K^{2n+1}  \\
&=&{\bf 1}_{j}
+\sum_{n=0}^{\infty}\frac{1}{(2n+2)!}\left(-\braket{z_{j}}\right)^{n}K^{2}
+\sum_{n=0}^{\infty}\frac{1}{(2n+1)!}\left(-\braket{z_{j}}\right)^{n}K  \\
&=&{\bf 1}_{j}
+\sum_{n=0}^{\infty}(-1)^{n}\frac{\left(\sqrt{\braket{z_{j}}}\right)^{2n}}
{(2n+2)!}K^{2}
+\sum_{n=0}^{\infty}(-1)^{n}\frac{\left(\sqrt{\braket{z_{j}}}\right)^{2n}}
{(2n+1)!}K  \\
&=&{\bf 1}_{j}
-\frac{1}{\braket{z_{j}}}\sum_{n=0}^{\infty}(-1)^{n+1}
\frac{\left(\sqrt{\braket{z_{j}}}\right)^{2n+2}}{(2n+2)!}K^{2}
+\frac{1}{\sqrt{\braket{z_{j}}}}\sum_{n=0}^{\infty}(-1)^{n}
\frac{\left(\sqrt{\braket{z_{j}}}\right)^{2n+1}}{(2n+1)!}K  \\
&=&{\bf 1}_{j}
-\frac{1}{\braket{z_{j}}}\sum_{n=1}^{\infty}(-1)^{n}
\frac{\left(\sqrt{\braket{z_{j}}}\right)^{2n}}{(2n)!}K^{2}
+\frac{1}{\sqrt{\braket{z_{j}}}}\sum_{n=0}^{\infty}(-1)^{n}
\frac{\left(\sqrt{\braket{z_{j}}}\right)^{2n+1}}{(2n+1)!}K  \\
&=&{\bf 1}_{j}
-\frac{1}{\braket{z_{j}}}\left(\cos(\sqrt{\braket{z_{j}}})-1\right)K^{2}
+\frac{\sin(\sqrt{\braket{z_{j}}})}{\sqrt{\braket{z_{j}}}}K  \\
&=&{\bf 1}_{j}
+\left(1-\cos(\sqrt{\braket{z_{j}}})\right)\frac{1}{\braket{z_{j}}}K^{2}
+\sin(\sqrt{\braket{z_{j}}})\frac{1}{\sqrt{\braket{z_{j}}}}K.
\end{eqnarray*}

If we define a normalized vector as
\[
\ket{\tilde{z}_{j}}=\frac{1}{\sqrt{\braket{z_{j}}}}\ket{z_{j}}
\ \Longrightarrow\ \braket{\tilde{z}_{j}}=1
\]
then 
\[
\frac{1}{\sqrt{\braket{z_{j}}}}K
=
\left(
\begin{array}{cc}
{\bf 0}_{j-1} & \ket{\tilde{z}_{j}}   \\
-\bra{\tilde{z}_{j}} & 0 
\end{array}
\right),\quad 
\frac{1}{\braket{z_{j}}}K^{2}
=
\left(
\begin{array}{cc}
-\ket{\tilde{z}_{j}}\bra{\tilde{z}_{j}} &  \\
  & -1
\end{array}
\right).
\]
Therefore
\begin{equation}
\mathrm{e}^{K}
=
\left(
\begin{array}{cc}
{\bf 1}_{j-1}-\left(1-\cos(\sqrt{\braket{z_{j}}})\right)
\ket{\tilde{z}_{j}}\bra{\tilde{z}_{j}}& 
\sin(\sqrt{\braket{z_{j}}})\ket{\tilde{z}_{j}}   \\
-\sin(\sqrt{\braket{z_{j}}})\bra{\tilde{z}_{j}} & 
\cos(\sqrt{\braket{z_{j}}})
\end{array}
\right).
\end{equation}
As a result we obtain the formula (\ref{eq:fundamental})
\begin{equation}
\mathrm{e}^{X_{j}}
=
\left(
\begin{array}{ccc}
{\bf 1}_{j-1}-\left(1-\cos(\sqrt{\braket{z_{j}}})\right)
\ket{\tilde{z}_{j}}\bra{\tilde{z}_{j}}& 
\sin(\sqrt{\braket{z_{j}}})\ket{\tilde{z}_{j}} &  \\
-\sin(\sqrt{\braket{z_{j}}})\bra{\tilde{z}_{j}} & 
\cos(\sqrt{\braket{z_{j}}}) & \\
  &  &  {\bf 1}_{n-j}
\end{array}
\right).
\end{equation}
%


\end{document}